\documentclass{elsart}

\usepackage{amssymb}

\newcommand{\beq}{\begin{equation}}
\newcommand{\eeq}{\end{equation}}
\newcommand{\md}{{\rm d}}
\newcommand{\imu}{{\rm i}}
\newcommand{\mrm}[1]{\mathrm{#1}}
\newcommand{\mcal}[1]{\mathcal{#1}}
\newcommand{\re}{\mathop{\rm Re}}
\newcommand{\im}{\mathop{\rm Im}}

\begin{document}

\begin{frontmatter}

\title{Quasi-Hermiticity in infinite-dimensional Hilbert spaces}

\author[RK]{R. Kretschmer\corauthref{cor}},
\corauth[cor]{Corresponding author.}
\ead{kretschm@hepth2.physik.uni-siegen.de}
\author[LS]{L. Szymanowski}
\ead{lech.szymanowski@fuw.edu.pl}

\address[RK]{Fachbereich Physik, Universit\"at Siegen, 
Emmy-Noether-Campus, 57068 Siegen, Germany}

\address[LS]{Soltan Institute for Nuclear Studies, Hoza 69, 
00681 Warsaw, Poland}

\begin{abstract}
In infinite-dimensional Hilbert spaces, the application of the concept
of quasi-Hermiticity to the description of non-Hermitian Hamiltonians
with real spectra may lead to problems related to the definition of 
the metric operator. We discuss these problems by examining some 
examples taken from the recent literature and propose a formulation 
that is free of these difficulties.
\end{abstract}


\end{frontmatter}

\section{Introduction}
\label{s1}

In the past few years, the study of non-Hermitian Hamiltonians has
attracted much interest, because under certain conditions, 
non-Hermitian Hamiltonians may have a real spectrum and therefore may
describe realistic physical systems \cite{Bender}. Many examples of 
this kind are known \cite{Bender,Cannata}. Despite that, the physical
interpretation of quantum theories based on non-Hermitian Hamiltonians 
remains obscure.

Recently, the concepts of quasi- or pseudo-Hermitian operators 
\cite{Scholtz,Mostafazadeh1} have become very popular in the attempts 
to overcome the problems to find a physical interpretation for such 
theories.

These concepts seem to be adequate to describe systems whose 
underlying Hilbert space is finite-dimensional. In this paper, we want 
to point out that they have to be used with some care when applied to 
the more interesting theories that are defined in infinite-dimensional 
Hilbert spaces. We show that in many examples that have been discussed 
in the literature, unbounded metric operators appear, and emphasize 
that this is incompatible with the concepts of quasi- or 
pseudo-Hermiticity. Then we demonstrate how these difficulties can be 
avoided within standard quantum mechanics based on Hermitian operators 
\cite{Kretschmer}.

\section{Quasi- and pseudo-Hermiticity}
\label{s2}

Given a Hilbert space $\mcal{H}$ with scalar product $(.,.)$, an 
operator $A: \mcal{H} \to \mcal{H}$ is called quasi-Hermitian, if 
there exists an operator $\eta: \mcal{D}_\eta \to \mcal{H}$ that has 
the following properties:
\begin{itemize}
\item the domain of definition of $\eta$ is the entire 
space, $\mcal{D}_\eta = \mcal{H}$,
\item the operator $\eta$ is Hermitian, $\eta^\dagger = \eta$,
\item the operator $\eta$ is positive definite, 
$(\varphi, \eta \varphi) > 0$ for all $\varphi \in \mcal{H}$, 
$\varphi \neq 0$,
\item $\eta$ is bounded, i.~e.\ for all $\varphi \in \mcal{H}$ there
exists a real, positive $k$ such that 
$\|\eta \varphi\| \leq k \|\varphi\|$,
\item $\eta A = A^\dagger \eta$.
\end{itemize}

This definition is taken from the work of Scholtz et al.\
\cite{Scholtz}, who give a very thorough discussion of quantum 
theories based on quasi-Hermitian operators. The importance of this 
concept lies in the fact that if one introduces a new scalar product 
with the metric operator $\eta$, such that for arbitrary $\varphi$, 
$\psi \in \mcal{H}$
\beq\label{2a}
(\varphi, \psi)_\eta := (\varphi, \eta \psi) \quad,
\eeq
thereby defining a new Hilbert space $\mcal{H}_\eta$, then the 
operator $A$ (which may be non-Hermitian with respect to the original
scalar product $(.,.)$) is Hermitian with respect to the new one:
\beq
(\varphi, A \psi)_\eta = (\varphi, \eta A \psi) 
= (\varphi, A^\dagger \eta \psi) = (A \varphi, \eta \psi) 
= (A \varphi, \psi)_\eta \quad.
\eeq
In this way, the scalar product $(.,.)_\eta$ can serve as the basis of
a quantum theory. 

The two conditions on the domain of definition and boundedness of
$\eta$ are not independent: According to the theorem of Hellinger and
Toeplitz \cite{Prugovecki}, any Hermitian operator that is defined on
the entire Hilbert space $\mcal{H}$ is bounded.

A notion closely related to quasi-Hermiticity is pseudo-Hermiticity. 
Its importance in the current discussion of non-Hermitian Hamiltonians 
with real spectra has been emphasized in the work of Mostafazadeh 
\cite{Mostafazadeh1}. In \cite{Mostafazadeh1}, an operator $A$ is 
called pseudo-Hermitian, if a Hermitian automorphism $\tilde{\eta}$ 
exists that fulfills $A^\dagger = \tilde{\eta} A \tilde{\eta}^{- 1}$. 
Being an automorphism, its domain of definition is the entire space, 
so that (again by virtue of the theorem of Hellinger and Toeplitz) it 
is bounded. On the other hand, as shown in \cite{Scholtz}, the $\eta$ 
appearing in the definition of quasi-Hermiticity is an automorphism, 
too. Thus quasi-Hermiticity and pseudo-Hermiticity are identical 
except for the requirement that, contrary to $\tilde{\eta}$, $\eta$ 
has to be positive-definite \cite{Mostafazadeh2}. This 
positive-definiteness of $\eta$ ensures the positive-definiteness
of the scalar product (\ref{2a}), and is thus a necessary 
requirement if one attempts to construct a Hilbert space based on the 
scalar product (\ref{2a}).

The distinction between quasi- and pseudo-Hermiticity is not always
made, see e.~g. \cite{Mostafazadeh1,Ahmed1,Solombrino,Ahmed2,Fityo}. 
In our subsequent analysis, we will consider only systems with 
quasi-Hermitian Hamiltonians.

In finite-dimensional Hilbert spaces $\mcal{H}$ the condition on the 
domain of definition of $\eta$ can be easily fulfilled and the 
boundedness condition always holds. This is the reason why 
quasi-Hermiticity is so useful in this case. However, in 
infinite-dimensional Hilbert spaces, the domain of definition of 
$\eta$ and the boundedness of $\eta$ are important constraints. 

If the domain of definition of $\eta$ is smaller than the space 
$\mcal{H}$, so that, say, $\varphi$ is not in the domain of definition 
of $\eta$, whereas $\psi$ is, then $(\varphi, \eta \psi)$ is 
well-defined, but $(\eta \varphi, \psi)$ is not defined. Therefore, 
although $\eta$ may be Hermitian, the equation
\beq\label{2d}
(\varphi, \psi)_\eta = (\psi, \varphi)_\eta^*
\eeq
does not always make sense. But (\ref{2d}) is one of the fundamental
defining relations of a scalar product.

The boundedness of $\eta$ is important, because Hilbert spaces are by
definition norm complete. This means that they contain all limits of
Cauchy sequences, i.~e.\ sequences $\xi_1, \xi_2, \ldots$ of vectors 
with the property that for all $\varepsilon > 0$ there exists a 
positive number $M(\varepsilon)$ such that
\beq\label{2b}
\|\xi_n - \xi_m\| < \varepsilon \quad \mbox{for all } 
n, m > M(\varepsilon) \quad.
\eeq
Now the norm in (\ref{2b}) explicitly depends on the scalar product 
chosen in the Hilbert space, $\|\xi_n\| = \sqrt{(\xi_n, \xi_n)}$. A 
change of the scalar product as in (\ref{2a}) may change the 
convergence properties of sequences. The requirement of boundedness of 
$\eta$ just ensures that this does not happen: Given a Cauchy sequence 
in the norm implied by $(.,.)$, one finds for the norm 
$\|\xi_n\|_\eta = \sqrt{(\xi_n, \xi_n)_\eta}$ implied by the scalar 
product $(.,.)_\eta$:
\[
\|\xi_n - \xi_m\|_\eta = \sqrt{(\xi_n - \xi_m, \eta(\xi_n - \xi_m))}
\leq \sqrt{\|\xi_n - \xi_m\| \|\eta(\xi_n - \xi_m)\|} \quad.
\]
Here we have used the Cauchy-Schwarz inequality. Now $\eta$ is
bounded, $\|\eta(\xi_n - \xi_m)\| \leq k \|\xi_n - \xi_m\|$ for some 
$k$, so that
\[
\|\xi_n - \xi_m\|_\eta \leq \sqrt{k} \, \|\xi_n - \xi_m\| 
\quad.
\]
Thus $\xi_1, \xi_2, \ldots$ is a Cauchy sequence with respect to the
norm $\|.\|_\eta$ if it is a Cauchy sequence with respect to the norm 
$\|.\|$. In \cite{Scholtz} it is shown that the converse statement is 
also true. Therefore, the Hilbert spaces $\mcal{H}$ and 
$\mcal{H}_\eta$ {\em contain the same vectors}.

As we will discuss in the next section, many examples of metric
operators $\eta$ found in the literature are {\em not\/} bounded. This
complicates the situation considerably. Let us first illustrate this 
with a very simple example: Consider an infinite set 
$\psi_1, \psi_2, \ldots$ of orthonormal vectors in some Hilbert space 
$\mcal{H}$,
\[
(\psi_n, \psi_m) = \delta_{nm} \quad \mbox{for all } n, m \quad.
\]
Within $\mcal{H}$, define the infinite sequence $\xi_1, \xi_2, \ldots$ 
with
\beq\label{2c}
\xi_n = {\psi_n \over \sqrt{n}} \quad \mbox{ for all } n \quad.
\eeq
Since 
$\|\xi_n - \xi_m\| = \sqrt{1 / n + 1 / m}$, this sequence is a Cauchy
sequence, thus its limit in $\mcal{H}$ exists,
\[
\lim_{n \to \infty} \xi_n \in \mcal{H} 
\]
(actually $\lim_{n \to \infty} \xi_n = 0$ in $\mcal{H}$). Now consider 
the unbounded linear operator $\eta$ defined by
\beq
\eta \psi_n := n \psi_n \quad \mbox{for all } n 
\eeq
and the Hilbert space $\mcal{H}_\eta$ with the new scalar product 
$(\varphi, \psi)_\eta = (\varphi, \eta \psi)$. For the sequence 
(\ref{2c}) one now finds $\|\xi_n - \xi_m\|_\eta = \sqrt{2}$ for all
$n \neq m$, so that the limit $\lim_{n \to \infty} \xi_n$ does not 
exist in $\mcal{H}_\eta$. In other words, the Hilbert spaces 
$\mcal{H}$ and $\mcal{H}_\eta$ {\em consist of different vectors}.

\section{Examples}
\label{s3}

In this section we want to discuss various examples of $\eta$ 
operators taken from the recent literature.

In \cite{Ahmed1} (see also \cite{Solombrino}) the positive-definite
operator 
\beq\label{3d}
\eta = \e^{- \theta p} \quad,
\eeq
where $p$ is the momentum operator and $\theta$ is a real number, is 
used to show that the complex Morse potential
\beq\label{3c}
V(x) = (A + \imu B)^2 \e^{- 2 x} - (2 C + 1) (A + \imu B) \e^{- x}
\eeq
($A$, $B$ and $C$ being real) is quasi-Hermitian. Indeed, for
$\theta = 2 \arctan(B / A)$, one obtains
\[
\e^{- \theta p} V(x) \e^{\theta p} = V(x + \imu \theta) 
= (V(x))^\dagger 
\quad.
\]
But (\ref{3d}) is not an automorphism in the space 
$L_2(- \infty, \infty)$ of square-integrable functions. If one defines
\beq
(\varphi, \psi)_\eta := (\varphi, \eta \psi)_{L_2} 
\equiv \int \limits_{- \infty}^\infty \md x \, \varphi^*(x) 
(\eta \psi)(x)
\eeq
as a new scalar product {\em for all\/} $\varphi$, $\psi \in L_2$, one
immediately faces inconsistencies: Take
\[
\varphi(x) = \exp( x - \e^x) \in L_2 \quad,\quad
\psi(x) = \e^{- x^2} \in L_2 \quad \mbox{and} \quad
\eta = \e^{- \pi p} \quad.
\]
The matrix element
\[
(\varphi, \eta \psi)_{L_2} 
= \int \limits_{- \infty}^\infty \md x \, \e^{x - \e^{x}} 
\e^{- x^2 - 2 \imu \pi x + \pi^2}
\]
is well-defined since the integrand vanishes for $x \to \pm \infty$ at
least exponentially. But
\[
(\eta \varphi, \psi)_{L_2}
= - \int \limits_{- \infty}^\infty \md x \, \e^{x + \e^x} \e^{- x^2} 
\]
is not defined, since the integrand vanishes for $x \to - \infty$, but
diverges for $x \to \infty$. The reason is (as in (\ref{2d})) that 
although $\varphi \in L_2$, the function 
$(\eta \varphi)(x) = \varphi(x + \imu \pi) = - \exp(x + \e^x)$ is not 
in $L_2$. Thus, despite the ``Hermitian appearance'' of $\eta$, the 
statement $(\varphi, \psi)_\eta^* = (\psi, \varphi)_\eta$ does not 
hold for all square-integrable functions.

Another example is provided by Hamiltonians of the form
\beq\label{3a}
H = {(p - \phi(x))^2 \over 2 m} + V(x)
\eeq
where $\phi(x)$ is a complex function and $V(x)$ is real. Similar 
models are investigated in \cite{Ahmed2,Mostafazadeh3,Bagchi}. A 
special case of (\ref{3a}) is the model of Hatano and Nelson 
\cite{Hatano}, which is obtained for 
$\phi(x) = - \imu g = \mrm{const}$. For (\ref{3a}) the 
positive-definite metric operator
\[
\eta = \exp \left( 2 \int\limits_{x_0}^x \md y \, \im \phi(y) \right)
\]
with arbitrary $x_0$ can be chosen (see 
\cite{Ahmed2,Mostafazadeh3,Bagchi}). Depending on the choice of 
$\phi(x)$, this operator may be unbounded \cite{Ahmed2}.

As a last example, we refer to \cite{Bagchi}, where among other models 
the case $H = p^2 + V(x)$,
\beq\label{3b}
V(x) = - g^2(x) + k - \imu {\md g \over \md x} \quad,\quad
\eta =  g(x) - \imu {\md \over \md x} 
\eeq
with real $g(x)$ and $k$ is investigated. The potential in (\ref{3b}) 
is related to supersymmetric quantum mechanics. The operator $\eta$ in 
(\ref{3b}) is an example of a metric operator that is not 
positive-definite. It is also not bounded; even for well-behaved 
$g(x)$, the derivative will spoil the boundedness.

\section{Construction without metric operator}
\label{s4}

In the previous sections we have shown that an unbounded metric 
operator $\eta$ cannot be used to define a consistent Hilbert space
structure. Now we give an alternative construction which is not based 
on the introduction of an $\eta$ operator as in (\ref{2a}) from the 
very beginning.

Consider the following situation: We are given a (non-Hermitian)
Hamiltonian $H$, an infinite, discrete set of eigenvectors $\psi_n$ of
$H$ that are elements of a Hilbert space $\tilde{\mcal{H}}$ (endowed 
with the scalar product $(.,.)_{\tilde{\mcal{H}}}$) and have real 
eigenvalues $E_n$.

The space $\tilde{\mcal{H}}$ may be the space $L_2(- \infty, \infty)$,
but in general this will not be the case. We emphasize that we do not 
assume any form of completeness of the $\psi_n$ such as, e.~g., the 
existence of a complete biorthonormal set of eigenvectors 
\cite{Mostafazadeh1}. Assumptions like this are often made, but to our 
knowledge, in the examples of non-Hermitian Hamiltonians with real
spectra treated in the literature, their validity is not examined.

We start the construction by considering the vector space $\mcal{V}$ 
that is defined as the span (the set of finite superpositions) of the 
vectors $\psi_1, \psi_2, \ldots$ We can define a scalar product 
$(.,.)_\mcal{V}$ in $\mcal{V}$ that fulfills
\beq
(\psi_n, \psi_m)_\mcal{V} = \delta_{nm} \quad \mbox{for all } n, m
\eeq
(possibly of the form 
$(\psi_n, \psi_m)_\mcal{V} 
= (\psi_n, \eta \psi_m)_{\tilde{\mcal{H}}}$). The completion of 
$\mcal{V}$ with respect to its norm (i.~e.\ the combination of 
$\mcal{V}$ with all limits of Cauchy sequences of vectors in 
$\mcal{V}$) yields a separable Hilbert space $\mcal{H}$
\cite{Prugovecki,Kretschmer,Mostafazadeh4}. In this space
\begin{itemize}
\item the set $\{\psi_1, \psi_2, \ldots\}$ is a complete orthonormal
system of vectors and
\item the Hamiltonian $H$ (more precisely the closed extension of the
restriction of $H$ to $\mcal{V}$) is Hermitian.
\end{itemize}
The last property can be easily seen by noting that all vectors in
$\mcal{H}$ are (possibly infinite) superpositions of the eigenvectors
$\psi_n$, e.~g.\ $\varphi = \sum_n c_n \psi_n$, 
$\psi = \sum_n \tilde{c}_n \psi_n$, thus
\[
(\varphi, H \psi)_\mcal{H} 
= \sum_{n, m} c_n^* \tilde{c}_m (\psi_n, H \psi_m)_\mcal{H} 
= \sum_n c_n^* \tilde{c}_n E_n
\]
and $(H \varphi, \psi)_\mcal{H}$ gives the same result, provided both
$\varphi$ and $\psi$ are in the domain of definition of $H$. Owing to 
these properties of $\mcal{H}$, this space can be used for a 
consistent quantum-mechanical formulation.

Since $\mcal{H}$ is an infinite-dimensional, separable Hilbert space, 
it is unitarily equivalent to any other infinite-dimensional, 
separable Hilbert space \cite{Prugovecki}, in particular to the space 
$L_2(- \infty, \infty)$. This means that an isomorphism 
$T: \mcal{H} \to L_2(- \infty, \infty)$ {\em must\/} exist with
\beq\label{4a}
(\varphi, \psi)_\mcal{H} = (T \varphi, T \psi)_{L_2} \quad
\mbox{for all } \varphi, \psi \in \mcal{H} \quad.
\eeq
With the help of the transformation $T$, one can define the 
Hamiltonian $\hat{H} = T H T^{- 1}$ that maps from $L_2$ to $L_2$, and 
its eigenvectors $\hat{\psi}_n = T \psi_n \in L_2$. The operator 
$\hat{H}$ is Hermitian in $L_2$: For $\hat{\varphi}$, $\hat{\psi}$ in 
the domain of definition of $\hat{H}$ one has
\begin{eqnarray*}
(\hat{\varphi}, \hat{H} \hat{\psi})_{L_2}
& = & (T^{- 1} \hat{\varphi}, T^{- 1} \hat{H} \hat{\psi})_\mcal{H}
= (T^{- 1} \hat{\varphi}, H T^{- 1} \hat{\psi})_\mcal{H}
= (H T^{- 1} \hat{\varphi}, T^{- 1} \hat{\psi})_\mcal{H} \\
& = & (T^{- 1} \hat{H} \hat{\varphi}, T^{- 1} \hat{\psi})_\mcal{H}
= (\hat{H} \hat{\varphi}, \hat{\psi})_{L_2} \quad.
\end{eqnarray*}
This is just a consequence of the unitary equivalence of the spaces
$\mcal{H}$ and $L_2(- \infty, \infty)$; it is merely a matter of taste 
whether the theory is formulated in $\mcal{H}$ or 
$L_2(- \infty, \infty)$.

This construction is very general, so that not every possible
transformation $T$ can be expected to be physically meaningful. See
\cite{Kretschmer} for a discussion of this aspect. Note that
\begin{itemize}
\item any reference to a metric operator $\eta$ has disappeared from
the construction; we are only using the reality of the spectrum of
$H$.
\item still, in cases in which it is possible to talk about the
Hermitian adjoint of $T$, one has 
$(\varphi, \psi)_\mcal{H} = (\varphi, T^\dagger T \psi)_{L_2}$, which 
looks like $\eta = T^\dagger T$. In fact, decompositions of $\eta$ 
like this are often used, because they guarantee the positive 
semi-definiteness of $\eta$ \cite{Mostafazadeh1,Solombrino,Fityo}.
\end{itemize}

Let us apply this construction to the examples mentioned in
Section~\ref{s3}: For the complex Morse potential (\ref{3c}), the
transformation
\[
T = \e^{- \theta p / 2}
\]
renders the potential Hermitian in the space $L_2(- \infty, \infty)$:
\beq
\hat{V} = T V T^{- 1} 
= (A^2 + B^2) \e^{- 2 x} - (2 C + 1) \sqrt{A^2 + B^2} \, \e^{- x} 
= \hat{V}^\dagger \quad.
\eeq
The Schr\"odinger equation for this {\em real\/} Morse potential has
the usual, well-known eigenfunctions 
$\hat{\psi}_n \in L_2(- \infty, \infty)$. Therefore, the functions
\[
\psi_n(x) = (T^{- 1} \hat{\psi}_n)(x) 
= \hat{\psi}_n(x - \imu \theta / 2) \in \mcal{H} \quad,
\]
which are not necessarily square-integrable, are the eigenfunctions of
the Schr\"o\-din\-ger equation for the complex Morse potential 
(\ref{3c}).

It is crucial to realize that $T$ is not something like the 
square-root of $\eta$ in (\ref{3d}) \cite{Mostafazadeh4}. It is a map 
from $\mcal{H}$ to $L_2$, whereas $\eta$ would have to be an 
automorphism $\tilde{\mcal{H}} \to \tilde{\mcal{H}}$. As such, $T$ is 
always bounded, $\|T\| = 1$.

The example (\ref{3a}) can also be handled easily: The transformation
\[
T = \exp \left( - \imu \int\limits_{x_0}^x \md y \, \phi(y) \right)
\]
gives
\beq
\hat{H} = T H T^{- 1} = {p^2 \over 2 m} + V(x) = \hat{H}^\dagger
\quad.
\eeq
This transformation can be factorized into $T = T_\mrm{g} T_\mrm{u}$,
with 
$T_\mrm{g} 
= \exp( - \imu \int_{x_0}^x \md y \, \allowbreak \re \phi(y))$ 
and $T_\mrm{u} = \exp(\int_{x_0}^x \md y \, \im \phi(y))$. Here the
first term $T_\mrm{g}$ is just a usual, unitary gauge transformation 
(thus an automorphism), its contribution cancels in (\ref{4a}). The 
second term $T_\mrm{u}$ is the non-trivial part that will in general 
map between different Hilbert spaces.

For the case of the supersymmetric model (\ref{3b}), we restrict the
discussion to $k = 0$. Then the Hamiltonian can be written in the
factorized form
\[
H = (p - g(x)) (p + g(x)) \quad.
\]
Defining $G(x) = \int_{x_0}^x \md y \, g(y)$, one can apply a gauge
transformation (note that $g(x)$ is a real function):
\[
\e^{\imu G(x)} H \e^{- \imu G(x)} = (p - 2 g(x)) p \quad.
\]
Multiplying this with $\sqrt{p}$ and $\sqrt{p}^{- 1}$ from the left 
and right, respectively, one obtains the operator
\begin{eqnarray}
\hat{H}
& = & \sqrt{p} \, \e^{\imu G(x)} H \e^{- \imu G(x)} \sqrt{p}\,^{- 1}
= p^2 - 2 \sqrt{p} \, g(x) \sqrt{p} \label{4b} \\
& = & p^2 - 2 p g(x + \imu / (2 p)) = p^2 - 2 g(x - \imu / (2 p)) p 
\quad. \label{4c}
\end{eqnarray}
The expressions (\ref{4c}) are valid for analytic $g(x)$. They can be 
derived by noting that 
$\sqrt{p} \, x \sqrt{p} = p (x + \imu / (2 p))$, which generalizes by 
induction to $\sqrt{p} \, x^n \sqrt{p} = p (x + \imu / (2 p))^n$. The 
relations (\ref{4c}) show that $\hat{H}$ is Hermitian in $L_2$. 
Therefore, the transformation $T: \mcal{H} \to L_2$ can be chosen to 
be
\[
T = \sqrt{p} \, \e^{\imu G(x)} \quad.
\]
Due to the appearance of $\sqrt{p}$ and $1 / p$ in (\ref{4b}) and
(\ref{4c}), the Hamiltonian $\hat{H}$ may not be well-defined. One 
may, however, attempt to solve the Schr\"odinger equation in the
representation in which $p$ is diagonal. It should then be possible to
give a well-defined meaning to (\ref{4b}) or (\ref{4c}). But the
details of such a construction depend on the choice of $g(x)$ in
(\ref{3b}) and go beyond the scope of the present paper.

Equation (\ref{4b}) provides an example in which the Hermitian
Hamiltonian $\hat{H}$ assumes a rather unusual form that cannot be 
interpreted as a sum of kinetic and potential energy if the operators 
$x$ and $p$ have their usual meaning.

Let us emphasize that the construction outlined above can be applied 
in situations in which previous analyses have led to unbounded metric 
operators. Besides this, it is not necessary to assume any form of 
completeness for the eigenfunctions of the Hamiltonian. Therefore, 
this approach offers an alternative to the recent application of the 
theory of quasi-Hermitian operators to the study of non-Hermitian 
Hamiltonians with real spectra \cite{Mostafazadeh1}.

\section{Conclusions and outlook}
\label{s5}

In this paper we have demonstrated that the notion of 
quasi-Hermiticity cannot be used to define a consistent quantum 
theory if the requirement of the boundedness of the metric operator 
$\eta$ is not fulfilled. Some examples in which unbounded metric 
operators are used to describe theories with non-Hermitian 
Hamiltonians and real spectra have been discussed in detail. In order 
to map such theories in a consistent way to Hermitian theories, we 
have presented an alternative formulation that is based only on the 
unitary equivalence of infinite-dimensional, separable Hilbert spaces. 
This unitary equivalence ensures the existence of a transformation $T$ 
that maps a given non-Hermitian Hamiltonian with real spectrum to a 
Hermitian one via a similarity transformation.

An interesting question that we have not addressed here concerns the
uniqueness of the transformation $T$. It has been shown in
\cite{Scholtz} that a (bounded) metric operator $\eta$ is only
uniquely defined if an appropriately chosen {\em set of observables\/}
is considered. The same is true for the transformation $T$. In
\cite{Kretschmer} some attempts to find such a set of observables have 
been discussed. In our opinion this aspect has not received enough
attention yet, but will be crucial for the understanding of theories 
with non-Hermitian Hamiltonians.

\section*{Acknowledgments}

We thank the referee for his very constructive remarks. The work of 
L.~S. is supported in part by the French-Polish scientific agreement 
Polonium.


\begin{thebibliography}{00}
\bibitem{Bender}
	C. M. Bender, S. Boettcher, Phys.\ Rev.\ Lett.\ \textbf{80}, 
	5243 (1998);
	C. M. Bender, S. Boettcher, P. Meisinger, J. Math.\ Phys.\
	\textbf{40}, 2201 (1999).
\bibitem{Cannata}
	F. Cannata, G. Junker, J. Trost, Phys.\ Lett.\ A 
	\textbf{246}, 219 (1998);
	M. Znojil, Phys.\ Lett.\ A \textbf{259}, 220 (1999);
	P. Dorey, C. Dunning, R. Tateo, J. Phys.\ A: Math.\ Gen.\
	\textbf{34}, L391 (2001).
\bibitem{Scholtz}
	F. G. Scholtz, H. B. Geyer, F. J. W. Hahne, Ann.\ Phys.\ NY
	\textbf{213}, 74 (1992).
\bibitem{Mostafazadeh1}
	A. Mostafazadeh, J. Math.\ Phys.\ \textbf{43}, 205 (2002);
	A. Mostafazadeh, J. Math.\ Phys.\ \textbf{43}, 2814 (2002);
	A. Mostafazadeh, J. Math.\ Phys.\ \textbf{43}, 3944 (2002).
\bibitem{Kretschmer}
	R. Kretschmer, L. Szymanowski, Preprint quant-ph/0105054
	(2001), unpublished.
\bibitem{Prugovecki}
	E. Prugove\v{c}ki, \textit{Quantum Mechanics in Hilbert
	Space\/}, Academic Press (New York, 1981).
\bibitem{Mostafazadeh2} 
	A. Mostafazadeh, J. Math.\ Phys.\ \textbf{44}, 974 (2003);
	A. Mostafazadeh, Preprint quant-ph/0308028 (2003).
\bibitem{Ahmed1}
	Z. Ahmed, Phys.\ Lett.\ A \textbf{290}, 19 (2001).
\bibitem{Solombrino} 
	L. Solombrino, J. Math.\ Phys.\ \textbf{43}, 5439 (2002).
\bibitem{Ahmed2}
	Z. Ahmed, Phys.\ Lett.\ A \textbf{294}, 287 (2002). 
\bibitem{Fityo}
	T. V. Fityo, J. Phys.\ A: Math.\ Gen.\ \textbf{35}, 5893
	(2002).
\bibitem{Mostafazadeh3}
	A. Mostafazadeh, Mod.\ Phys.\ Lett.\ A \textbf{17}, 1973 
	(2002).
\bibitem{Bagchi}
	B. Bagchi, C. Quesne, Phys.\ Lett.\ A \textbf{301}, 173 
	(2002).
\bibitem{Hatano}
	N. Hatano, D. R. Nelson, Phys.\ Rev.\ Lett.\ \textbf{77}, 570
	(1996);
	N. Hatano, D. R. Nelson, Phys.\ Rev.\ B \textbf{56}, 8651 
	(1997).
\bibitem{Mostafazadeh4}
	A. Mostafazadeh, J. Phys.\ A: Math.\ Gen.\ \textbf{36}, 7081 
	(2003). 
\end{thebibliography}
\end{document}